\documentclass[mathleft]{an}
\usepackage{graphicx}
\usepackage{graphicx,epstopdf}
\usepackage{amsmath}
\usepackage{amssymb}
\usepackage{url}
\usepackage{txfonts}
\usepackage{times}
\usepackage{natbib}
\bibpunct{(}{)}{;}{a}{}{,} 
\def\hu{HU\,Aqr}

\begin{document}

\Pagespan{000}{}
\Yearpublication{0000}%
\Yearsubmission{0000}%
\Month{00}%
\Volume{000}%
\Issue{00}%

\title{On the ephemeris of the eclipsing polar HU Aquarii. II:\\ New eclipse epochs obtained 2014 -- 2018}

\author{A.D.~Schwope\inst{1}\fnmsep\thanks{Corresponding author:
  \email{aschwope@aip.de}\newline}
\and
B.D.~Thinius\inst{2}
}
\titlerunning{HU Aqr: New eclipse epochs obtained 2014 -- 2018}
\authorrunning{A.D.~Schwope \& B.~Thinius}
\institute{Leibniz-Institut f\"ur Astrophysik Potsdam (AIP),
              An der Sternwarte 16, 14482 Potsdam, Germany
              \and
Inastars Observatory, Hermann-Struve-Str. 10, 14469 Potsdam, Germany
}

\received{}
\accepted{}
\publonline{}

\abstract{The magnetic cataclysmic variable HU Aquarii displayed pronounced
  modulations of its eclipse timing. These were intensively modeled and
  discussed in recent years in the framework of planets orbiting the binary or
  the Applegate effect. No scenario yielded a unique and satisfactory
  interpretation of the data. Here we present 26 new eclipse epochs obtained
  between 2014 and 2018. The steep and continuous decrease of the orbital
  period observed in the time interval 2010 - 2013 has slowed down sometimes
  before mid 2016. The new slope in the $(O-C)$-diagram of eclipse arrival
  times will further constrain physical models of its complex shape. 
}


\maketitle
%

\section{Introduction}
\hu\ is an eclipsing magnetic cataclysmic variable with a 125\,min orbital period. When discovered in 1993 as the optical counterpart to the soft X-ray and EUV sources RX\,J2107.9-0518/RE2107-05
\citep{schwope+93, hakala+93} it was the brightest eclipsing object displaying the most extended eclipse. Those properties triggered broad observational studies
to disentangle accretion phenomena and the accretion geometry in a strongly 
magnetic environment. Particular emphasis was given to model the detailed eclipse structure \cite[see e.g. ][]{schwope+01, vrielmann+schwope01}.

Comprehensive X-ray and EUV observations with the ROSAT and EUVE 
satellites took place  between 1992 and 1998 \citep{schwope+01}. 
These studies established the eclipse egress as a  fiducial mark to determine
the orbital period and a long-term ephemeris and was used by all researches since then \citep{vogel+08,schwarz+09,qian+11,godz+12,schwope+thinius14, bours+14,godz+15}. Already this early study from 2001 gave some evidence for deviations of the eclipse egress time from a linear relationship between the cycle counts and the time of arrival. The size of the effect, however, $\pm5$\,s, was still compatible with a migration of the accretion spot over the surface of the white dwarf.

\cite{schwarz+09} were the first to discuss the timing residuals, that were then larger than the size of the white dwarf,  in terms of an unseen third body and derived a possible mass of $M_3 = 5 M_{\rm Jup}$ for a planetary companion. Alternative explanations were also discussed, namely Applegate's mechanism \citep{applegate92} or an extra mechanism of angular momentum loss acting in HU Aqr.

\cite{qian+11} added a further 11 eclipse epochs and claimed the discovery of a circumbinary planetary system around the accreting binary. Usage of the p-word raised more interest in the system and triggered further theoretical and observational studies. While both the Qian et al.~data and their model were soon rejected by superior data obtained at similar epochs \citep{godz+12,bours+14,godz+15} and by stability considerations \citep[see e.g.~][]{horner+11,wittenmyer+12}, the scenario assuming a planetary system was nevertheless further intensively studied. 

The two-planet model proposed by \cite{godz+12} that was derived using data obtained prior to 2012 (up to cycle 78100) was ruled out by \cite{schwope+thinius14} (henceforth Paper I) based on data obtained in 2013 (cycles 86217 -- 86310). Their finding was confirmed by \cite{bours+14} and \cite{godz+15} who added more epochs with higher precision obtained from regular monitoring of the source with high-speed cameras. 

The most recent in-depth analysis of the $(O-C)$ residuals of the eclipse egress arrival times was presented by \cite{godz+15}. They conclude that the putative planetary system is either more complex than a co-planar two-planet configuration or that oscillations of the gravitational quadrupole moment of the donor star, the so-called Applegate mechanism in the formulation of \citet[][ and references therein]{lanza06} might be responsible for the large $(O-C)$ variations. All recent studies agree that more data are needed to properly distinguish between competing models. This short communication is meant to give an update on the evolution of the $(O-C)$ residuals based on data obtained with the same equipment as the previous communication by the same authors (Paper I).

Cycle counting follows the convention introduced by \cite{schwope+01} and phase zero refers to the first ROSAT observations with full phase coverage back in April 1993. 

\section{Observations and data reduction}
\hu\ was observed during 1 night in September 2014, ll nights in 2016 (September and November), 6 nights in 2017 (August, October and November), and 5 nights in July 2018. The equipment used for the observations is exactly the same as in Paper I, hence its description can be kept very short. 

All observations were conducted with the 14 inch Celestron reflector of
Schmidt Cassegrain type located at Inastars Observatory Potsdam (IAU MPC
observer code B15). The telscope is permanently installed at the roof of a
one-family dwelling in the suburb of Potsdam, Germany.  

All observations were done in white light. An ASTRONOMIK filter was inserted
to block strong emission lines at this light-polluted site.  
Individual images of the field of HU Aqr  were recorded with an SBIG ST-8XME
CCD as detector. The camera was always used with a 2x2 binning and always with
3 sec integration time on target. The time resolution achieved was 5.2 sec for
all runs described here. The start time of each exposure and the exposure
time was written into the fits headers. The computer equipment was correlated
with a time signal of atomic clocks every five minutes via the Network Time
Protocol. The measured time difference between servers on several continents
and the PC used to control the measurements was typically of the order of 10
ms or less and does not contribute to the error budget. 

The format of writing the start times of frames was changed from integer
(Paper I) to float (double precision) so that no systematic offset had to be
accounted for as in Paper I. A shutter latency of 0.77 s was found and was 
applied as an additional offset to individual timings of the eclipses. 

CCD data reduction followed standard procedures and included bias subtraction and
  flatfield correction. The analysis of the light curves, 
i.e.~differential photometry with respect to comparison star 'C' 
\citep{schwope+93}  was performed with AstroImageJ \citep{collins+17}.
The determination of eclipse egress times followed the scheme described in
Paper I. Approximate overall
brightness levels at the time of the observations were read from phase-folded
lightcurves. The quantity used was the flux ratio of the target with respect
to the comparison star (relative flux). 

A summary of the observations reported in this paper is given in
Table~\ref{t:log}, which lists the observation interval per night, the number
of frames obtained, and the maximum brightness (relative flux) prior  to the
eclipse to characterize the accretion state of HU Aqr. For comparison, the
maximum relative flux of our observations obtained in 2013 was 0.85 (see
Fig.~1 in Paper I). 

\begin{table}[t]
\caption{Time-resolved photometric observations of \hu\ obtained at Inastars
  Observatory in years 2014, 2016, 2017, and 2018. Given are the observation
  date, the time interval covered, and the number of frames obtained per
  night. The accretion state is encoded by giving the maximum  relative
  bright-phase flux prior to eclipse. A dash indicates that only parts of the
  binary cycle were covered and no unique information about the overall
  brightness could be extracted from the data. 
\label{t:log}}
\begin{tabular}{lcrr}
\hline
Date & Observation interval &  \# frames & state\\
\hline
20140927 & $ 2456928.278040 - .355020$ & 1292 & 0.35\\
20160905 & $ 2457637.428809 - .468838$ & 1106 & 0.24\\
20160906 & $ 2457638.295402 - .508805$ & 3140 & -"- \\
20160907 & $ 2457639.305958 - .391210$ & 1418 & -"- \\
20160908 & $ 2457640.287360 - .405364$ & 1860 & -"- \\
20160915 & $ 2457647.293444 - .346335$ &  883 & -"- \\
20160919 & $ 2457651.309938 - .361001$ &  852 & -"- \\
20160921 & $ 2457653.298070 - .368882$ & 1182 & -"- \\
20161127 & $ 2457720.175346 - .218559$ &  721 & -- \\
20161128 & $ 2457721.213724 - .248244$ &  576 & -- \\
20170828 & $ 2457994.354082 - .441516$ & 1477 & 0.66\\
20170829 & $ 2457995.302341 - .418170$ & 1955 & -"- \\
20171015 & $ 2458042.278304 - .420315$ & 2399 & 0.51\\
20171016 & $ 2458043.287405 - .373581$ & 1456 & -"- \\
20171117 & $ 2458075.188120 - .207617$ &  326 & --  \\
20171123 & $ 2458081.265321 - .283137$ &  297 & --  \\
20180723 & $ 2458324.354952 - .459001$ & 1729 & 0.28\\
20180724 & $ 2458323.361905 - .443743$ & 1366 & -"- \\
20180901 & $ 2458363.426169 - .440556$ &  241 & -- \\
20180903 & $ 2458365.329056 - .347401$ &  307 & -- \\
20181009 & $ 2458401.272701 - .310929$ &  637 & 0.35\\
\hline
\end{tabular}
\end{table}

\section{Analysis and results}
\subsection{New eclipse epochs}

\hu\ was encountered at intermediate accretion states during all occasions between 2014 and 2018. The brightness of the source as indicated by the maximum brightness in the last column of Tab.~\ref{t:log} was lower than observed in 2013 where it was fond at 0.85. Also the shape of the light curve was found to be variable from being double-humped in the high state to become single-humped at reduced brightness (i.e.~at reduced accretion rate). The pre-eclipse dip due to the intervening accretion stream was clearly present only in August and September 2017. The centre phase of the pre-eclipse dip was dependent on the brightness, it was centred 0.12 phase units before eclipse centre in August 2017 and 0.092 phase units before eclipse centre in September 2017.

This behaviour, the morphological changes of the light curves and the phasing of the pre-eclipse dip as a function of the mass accretion rate (brightness of the source) were presented already by \cite{schwope+01}, who show a collection of light curves in intermediate and high states (their Fig.~3) that appear to be similar to the new data.

The times of individual CCD frames were converted to dynamical time (TDB) in the form of Julian days and corrected to the barycentre of the Solar system using time utilities provided by \cite{eastman+10}\footnote{\url{http://astroutils.astronomy.ohio-state.edu/time/}}, to give what we refer to as BJD(TDB).

The times of individual eclipse egress were measured by averaging a few data points before and after the egress, computing the half-light intensity and reading the times with a cursor from a graph of the light curve (see Paper I for a graph illustrating the method). Uncertainties of individual eclipse timings were set to half the time resolution achieved, which is 2.6 s at all occasions. All new eclipse measurements are listed in Table~\ref{t:ecl}. In total there are 26 new eclipse epochs, an increase by about 10\%.

\begin{table}[t]
\caption{New mid egress times $t_{\rm e}$ and uncertainties $\Delta t_{\rm e}$ for the observed eclipses of HU Aqr in 2014 -- 2018 from Inastars Observatory. Times are given in BJD(TDB).
\label{t:ecl}
}
\begin{tabular}{rll}
\hline
Cycle & $t_{\rm e}$& $\Delta t_{\rm e}$\\
\hline
 90133 & 2456928.301931 & 0.000030  \\
 90134 & 2456928.388747 & 0.000030  \\
 98300 & 2457637.362839 & 0.000030  \\
 98301 & 2457637.449675 & 0.000030  \\
 98311 & 2457638.317802 & 0.000030  \\
 98312 & 2457638.404711 & 0.000030  \\
 98313 & 2457638.491519 & 0.000030  \\
 98323 & 2457639.359694 & 0.000030  \\
 98334 & 2457640.314718 & 0.000030  \\
 98461 & 2457651.340919 & 0.000030  \\
 98484 & 2457653.337738 & 0.000030  \\
 99254 & 2457720.189440 & 0.000030  \\
 99266 & 2457721.231257 & 0.000030  \\
102412 & 2457994.368037 & 0.000030  \\
102423 & 2457995.323057 & 0.000030  \\
102424 & 2457995.409877 & 0.000030  \\
102964 & 2458042.292862 & 0.000030  \\
102965 & 2458042.379698 & 0.000030  \\
102976 & 2458043.334682 & 0.000030  \\
103343 & 2458075.197739 & 0.000030  \\
103413 & 2458081.275169 & 0.000030  \\
106202 & 2458323.417057 & 0.000030  \\
106214 & 2458324.458869 & 0.000030  \\
106663 & 2458363.441204 & 0.000030  \\
106685 & 2458365.351258 & 0.000030  \\
107099 & 2458401.294880 & 0.000030  \\
\hline
\end{tabular}
\end{table}

\subsection{Eclipse timings from discovery papers}
A few more eclipse timings are reported in the literature but were 
not considered in the latest compilation of those events by
\citet{godz+15}. They can be found in the original discovery papers
\citep{hakala+93,schwope+93}. 
We converted the eclipse timings given in those papers to BJD(TDB). Although
the early data were obtained with rather low time resolution and have
corresponding large error bars their inclusion may turn out to be useful to
describe overall trends of the eclipse arrival times with respect to a chosen
model.  

Eclipse times given in the mentioned papers refer to the center of the eclipse
and were corrected to eclipse egress by adding half the eclipse length, which
is 291.7\,s for the optical and 292.8\,s for the RASS X-ray
data. \cite{hakala+93} did not give the original timings of the four eclipses
observed by them but they derive an ephemeris based on their data obtained
during three nights between May 28 and June 3, 1991, and we use the zero time
of their ephemeris. 

The recovered eclipse timings with the uncertainties as given in the
  original papers are listed in Tab.~\ref{t:early}. They enlarge the database
by further 10 data points that cover an extra 10,000 cycles of the binary, and extend
back to the ROSAT all-sky survey in 1990. 

\begin{table}[t]
\caption{Mid egress times $t_{\rm e}$ and uncertainties $\Delta t_{\rm e}$ for
  the eclipses of HU Aqr reported by \cite{schwope+93} (Ref. = 1) and
  \cite{hakala+93} (Ref. = 2) in 1990 and 1992. Times are given in BJD(TDB). 
\label{t:early}
}
\begin{tabular}{rlll}
\hline
Cycle & $t_{\rm e}$& $\Delta t_{\rm e}$& Ref.\\
\hline
 -10441 & 2448196.4276095 & 0.00081 & (1), RASS\\
  -3792 & 2448773.6968672 & 0.00005 & (2)\\
  -2377 & 2448896.5474982 & 0.00069 & (1)\\
  -2376 & 2448896.6346187 & 0.00029 & (1)\\  
  -2364 & 2448897.6762886 & 0.00029 & (1)\\
  -2354 & 2448898.5448385 & 0.00029 & (1)\\
  -2353 & 2448898.6314685 & 0.00029 & (1)\\
  -2342 & 2448899.5864684 & 0.00029 & (1)\\
  -2341 & 2448899.6734683 & 0.00029 & (1)\\
  -2181 & 2448913.5648568 & 0.00040 & (1)\\
\hline
\end{tabular}
\end{table}

\subsection{Eclipse ephemeris}
The newly determined times for the eclipse egress reported in tables
\ref{t:ecl} and \ref{t:early} were combined with those reported previously in
the literature \citep{schwope+01,schwarz+09,godz+12,bours+14,godz+15}. The
data obtained by \cite{qian+11} were not included because these were shown to
be offset from data obtained at similar epochs for an unknown reason
\citep[see the discussion in][]{godz+12}. However, the inclusion
or omission of those data does not change the overall appearance 
of the curve but hampers detailed modeling.

\begin{figure}[t]
\resizebox{\hsize}{!}{\includegraphics[clip=]{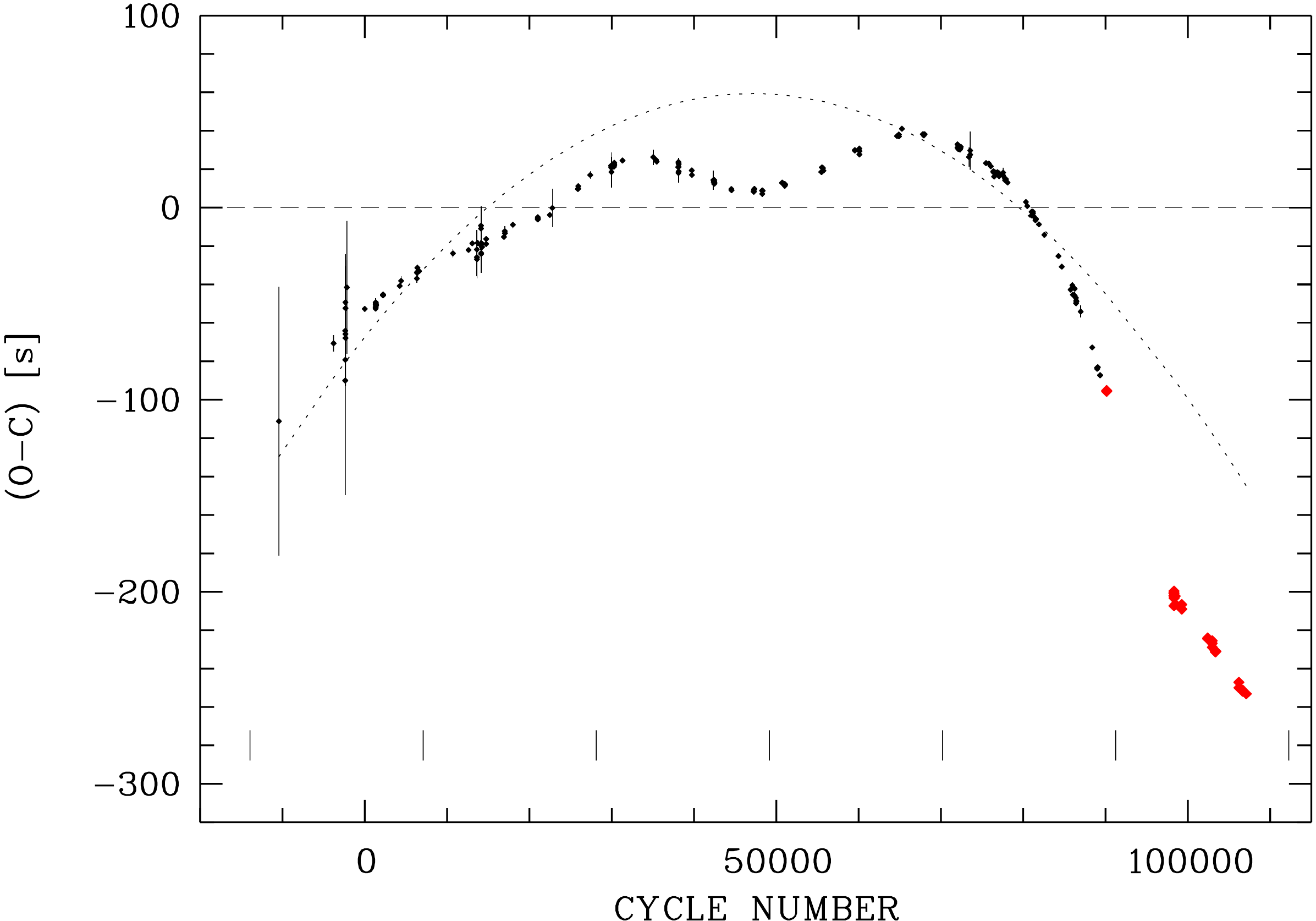}}
\caption{Observed minus calculated times of eclipse egress of \hu, 
according
  to equation 1. New original data from this work are shown with red
  rhombs. The dashed line indicates the improvements that are achieved through
  a quadratic fit (Eq.~2). Short vertical dashes at $(O-C) = -280$\,s indicate 5 year intervals
  beginning Jan 1, 1990.
}
\label{f:omc}
\end{figure}

The data set now comprises 244 individual eclipse timings, covering 28 years
and more than 115.000 orbital cycles of the $P_{\rm orb} = 125$\,min
binary. The set is composed of data obtained in the X-ray, the EUV,
the UV, and the optical regime. 
We take all reported measurement uncertainties at
face value thus ignoring the small
mismatches between the egress observed at optical and
e.g.~X-ray wavelengths. We also ignore a possible jitter that may occur due to
variable brightening of the accretion arc on the surface of the white
dwarf. Both effects are of order 3\,s or less. Including those would be
important for a detailed analysis of emission structures on the white dwarf or
for the development of detailed planetary model which is not attempted here
\citep[see the discussion in][who introduce a systematic uncertainty $\sigma$
  addressing those kinds of uncertainties]{godz+15}.  A weighted linear
regression to all data points yields the linear ephemeris for the eclipse
egress  
\begin{eqnarray}
{\rm BJD(TDB)} & = & 2449102.92061290(56) \nonumber \\ && + E \times 0.0868203923138(74)
\end{eqnarray}
(numbers in parenthesis give formal 1$\sigma$ uncertainties, 
reduced $\chi_\nu^2=20804$ for 242 d.o.f.). It is very obvious and well known
that a linear fit does not give a valid description of the data. The residuals
are however instructive and are shown in Fig.~\ref{f:omc}. 

The very first observation and main result of this short paper is that the 
accelerated decrease of the orbital period between 2010 and 2015 has slowed
down. The $(O-C)$-diagram displays a new slope that is obvious in the data
since year 2016.
The slope of the $(O-C)$-diagram seems to be constant since 
then with an ever decreasing orbital period. 

Any constant angular momentum loss will lead to a shorter orbital period and is to be decribed by a quadratic term in the $(O-C)$-diagram.
A quadratic fit of the form ${\rm BJD(TDB)} = T_0 + P E + 1/2 P \dot{P} E^2$
as a parameterization of such an extra, unspecified loss of angular momentum,  
reveals a better but still completely unsatisfactory representation of the data: 
\begin{eqnarray}
T_0 & = &
2449102.91983318(68) \nonumber\\
P & = & 0\fd086820454494(32)   \\
\dot{P} & = & (-1.518 \pm 0.008) \times 10^{-11}\nonumber
\end{eqnarray}
(reduced $\chi_\nu^2= 4087$ for 241 d.o.f.). 
Such a fit is illustrated with a dashed line in Fig.~\ref{f:omc}. With
respect to this quadratic fit, the $(O-C)$ values are even increasing after
cycle 98,000, also with an apparently constant slope. 

\section{Discussion}
The newest additions to the measured set of eclipse timings has revealed
another turn in the $(O-C)$-diagram of eclipse arrival times. The change
  in slope occurred between September 2014 and September 2016, where our data
  set has a gap. If one subtracts the 
linear term of the ephemeris according to Eq.~1 one would interpret the
behaviour of the $(O-C)$-values as if the decay of the orbital period would go
on but with reduced pace (Fig.~1). If a quadratic ephemeris is considered and
subtracted from the data, one gets the impression as if the decay of the orbit was
completely stopped prior to year 2016. A physical model
for a quadratic term, an extra amount of angular momentum loss, would need to
be found. Our measured rate of the period change is a factor 80 larger than
compatible with the secular change of the orbital period due to gravitational
radiation, $\dot{P}_{\rm GW} = -1.9 \times 10^{-13}$\,s/s \citep{bours+14}. 

The $(O-C)$ residuals give the impression as if some kind of
periodicity might be hidden in the data with a possible period of around
44,000 cycles (3900 days or roughly 10 years). Including a sine-curve in
  the fitting process does not help removing the large $(O-C)$ residuals, they 
  are not varying periodically. However, the amplitude of such an additional 
  sine-curve is of the order of 25\,s.

The deviations in the observed $(O-C)$ diagram of Fig.~1 are much larger than
the mentioned X-ray optical offsets and the phase jitter due to instationary
accretion.

\cite{godz+15} have intensively studied possible explanations of the
$(O-C)$-variations in the framework of Keplerian and N-body formulations of
the LTT effect (light travel time) for multiple planets and derive
evidence that the LTT hypothesis for the eclipse timing of \hu\ is unlikely.

\cite{voelschow+16} have studied the size of the Applegate effect for 11
post-common envelope binaries (PCEBs), among them \hu, and find that \hu\ is
one out of four systems whose amplitude of $(O-C)$-variation might still be
driven by an Applegate mechanism. Their assessment was based on the size of
the $(O-C)$ derived by \cite{godz+15} for their two-planet Keplerian model fit
with a quadratic ephemeris. This, as our fit in Eq.~2, has an uncomfortably
huge $\dot{P}$ of unknown origin. Furthermore the derived parameters were
found to be highly correlated and led \cite{godz+15} describing their
Keplerian model as essentially unconstrained. Nevertheless,
\cite{voelschow+16} used the best-fit parameters of the most influential
planet 'C' from the fit labeled JQ with a semi-amplitude $K_{\rm C} =
87.7$\,s, and period $P_{\rm C}=7101$\,days, to quantify the size of $(O-C)$
that an Applegate effect should be able to generate. Probably, the order of
magnitude of this effect is still valid. Would the periodicity as short as
10~years mentioned above an the size of the effect of order 25\,s instead of 87.7\,s this would be in favour of the applicability of the
mechanism in \hu, because the minimum energy required to drive this mechanism
scales with the inverse of the oscillation period. 

Whether or not just one explanation, a planetary system or the Applegate
mechanism in whatsoever form, or a combination therof \citep{bours+14}, or
some further not yet considered angular momentum loss mechanism is at work
here needs to be seen. Further regular monitoring of the source is of
pre-eminent importance. Whatever explanation is prefered eventually, it need
to give an explanation for the large apparent quadratic term in the
ephemeris and should also address the occurrence or non-occurrance of
$(O-C)$-variations in systems with very similar parameters as HU Aqr, for
example V808 Aur and UZ For, which also need long-term monitoring. 

\acknowledgement
We thank an anonymous referee for helpful comments.

\bibliographystyle{aa}
\bibliography{anhut}

\begin{thebibliography}{18}
\expandafter\ifx\csname natexlab\endcsname\relax\def\natexlab#1{#1}\fi

\bibitem[{{Applegate}(1992)}]{applegate92}
{Applegate}, J.~H. 1992, \apj, 385, 621

\bibitem[{{Bours} {et~al.}(2014){Bours}, {Marsh}, {Breedt}, {Copperwheat},
  {Dhillon}, {Leckngam}, {Littlefair}, {Parsons}, \& {Prasit}}]{bours+14}
{Bours}, M.~C.~P., {Marsh}, T.~R., {Breedt}, E., {et~al.} 2014, \mnras, 445,
  1924

\bibitem[{{Collins} {et~al.}(2017){Collins}, {Kielkopf}, {Stassun}, \&
  {Hessman}}]{collins+17}
{Collins}, K.~A., {Kielkopf}, J.~F., {Stassun}, K.~G., \& {Hessman}, F.~V.
  2017, \aj, 153, 77

\bibitem[{{Eastman} {et~al.}(2010){Eastman}, {Siverd}, \& {Gaudi}}]{eastman+10}
{Eastman}, J., {Siverd}, R., \& {Gaudi}, B.~S. 2010, \pasp, 122, 935

\bibitem[{{Go{\'z}dziewski} {et~al.}(2012){Go{\'z}dziewski}, {Nasiroglu},
  {S{\l}owikowska}, {Beuermann}, {Kanbach}, {Gauza}, {Maciejewski}, {Schwarz},
  {Schwope}, {Hinse}, {Haghighipour}, {Burwitz}, {S{\l}onina}, \&
  {Rau}}]{godz+12}
{Go{\'z}dziewski}, K., {Nasiroglu}, I., {S{\l}owikowska}, A., {et~al.} 2012,
  \mnras, 425, 930

\bibitem[{{Go{\'z}dziewski} {et~al.}(2015){Go{\'z}dziewski}, {S{\l}owikowska},
  {Dimitrov}, {Krzeszowski}, {{\.Z}ejmo}, {Kanbach}, {Burwitz}, {Rau},
  {Irawati}, {Richichi}, {Gawro{\'n}ski}, {Nowak}, {Nasiroglu}, \&
  {Kubicki}}]{godz+15}
{Go{\'z}dziewski}, K., {S{\l}owikowska}, A., {Dimitrov}, D., {et~al.} 2015,
  \mnras, 448, 1118

\bibitem[{{Hakala} {et~al.}(1993){Hakala}, {Watson}, {Vilhu}, {Hassall},
  {Kellett}, {Mason}, \& {Piirola}}]{hakala+93}
{Hakala}, P.~J., {Watson}, M.~G., {Vilhu}, O., {et~al.} 1993, \mnras, 263, 61

\bibitem[{{Horner} {et~al.}(2011){Horner}, {Marshall}, {Wittenmyer}, \&
  {Tinney}}]{horner+11}
{Horner}, J., {Marshall}, J.~P., {Wittenmyer}, R.~A., \& {Tinney}, C.~G. 2011,
  \mnras, 416, L11

\bibitem[{{Lanza}(2006)}]{lanza06}
{Lanza}, A.~F. 2006, \mnras, 369, 1773

\bibitem[{{Qian} {et~al.}(2011){Qian}, {Liu}, {Liao}, {Li}, {Zhu}, {Dai}, {He},
  {Zhao}, {Zhang}, \& {Li}}]{qian+11}
{Qian}, S.-B., {Liu}, L., {Liao}, W.-P., {et~al.} 2011, \mnras, 414, L16

\bibitem[{{Schwarz} {et~al.}(2009){Schwarz}, {Schwope}, {Vogel}, {Dhillon},
  {Marsh}, {Copperwheat}, {Littlefair}, \& {Kanbach}}]{schwarz+09}
{Schwarz}, R., {Schwope}, A.~D., {Vogel}, J., {et~al.} 2009, A\&A, 496, 833

\bibitem[{{Schwope} {et~al.}(2001){Schwope}, {Schwarz}, {Sirk}, \&
  {Howell}}]{schwope+01}
{Schwope}, A.~D., {Schwarz}, R., {Sirk}, M., \& {Howell}, S.~B. 2001, A\&A,
  375, 419

\bibitem[{{Schwope} \& {Thinius}(2014)}]{schwope+thinius14}
{Schwope}, A.~D. \& {Thinius}, B.~D. 2014, Astronomische Nachrichten, 335, 357

\bibitem[{{Schwope} {et~al.}(1993){Schwope}, {Thomas}, \&
  {Beuermann}}]{schwope+93}
{Schwope}, A.~D., {Thomas}, H.~C., \& {Beuermann}, K. 1993, A\&A, 271, L25

\bibitem[{{Vogel} {et~al.}(2008){Vogel}, {Schwope}, {Schwarz}, {Kanbach},
  {Dhillon}, \& {Marsh}}]{vogel+08}
{Vogel}, J., {Schwope}, A., {Schwarz}, R., {et~al.} 2008, in American Institute
  of Physics Conference Series, Vol. 984, High Time Resolution Astrophysics:
  The Universe at Sub-Second Timescales, ed. D.~{Phelan}, O.~{Ryan}, \&
  A.~{Shearer}, 264--267

\bibitem[{{V{\"o}lschow} {et~al.}(2016){V{\"o}lschow}, {Schleicher},
  {Perdelwitz}, \& {Banerjee}}]{voelschow+16}
{V{\"o}lschow}, M., {Schleicher}, D.~R.~G., {Perdelwitz}, V., \& {Banerjee}, R.
  2016, A\&A, 587, A34

\bibitem[{{Vrielmann} \& {Schwope}(2001)}]{vrielmann+schwope01}
{Vrielmann}, S. \& {Schwope}, A.~D. 2001, \mnras, 322, 269

\bibitem[{{Wittenmyer} {et~al.}(2012){Wittenmyer}, {Horner}, {Marshall},
  {Butters}, \& {Tinney}}]{wittenmyer+12}
{Wittenmyer}, R.~A., {Horner}, J., {Marshall}, J.~P., {Butters}, O.~W., \&
  {Tinney}, C.~G. 2012, \mnras, 419, 3258

\end{thebibliography}
  
\end{document}